\documentclass[useAMS,usenatbib]{mn2e}

\input{epsf}
\usepackage{graphicx}

\title[PKS 0040$-$005 and BAL Quasar Orientation]{Spectropolarimetry of PKS 0040$-$005 and the Orientation of Broad Absorption Line Quasars\footnotemark[0]\thanks{Based on observations collected at the European Southern Observatory, Paranal, project 71.B-0121(A).}}
 \author[M. S. Brotherton, C. De Breuck, and J. J. Schaefer]{M. S. Brotherton$^{1}$\thanks{E-mail: mbrother@uwyo.edu (MSB), ESO Visitor}, C. De Breuck$^{2}$\footnotemark[1]\thanks{E-mail: cdebreuc@eso.org.}, and J. J. Schaefer$^{3}$\footnotemark\thanks{E-mail: schaefjj@ufl.edu}\\ 
$^{1}$Department of Physics and Astronomy, University of Wyoming, Laramie, WY, 82072, USA\\
$^{2}$European Southern Observatory, Karl Schwarzschild Strasse 2, Garching Bei Munchen, 85748, Germany\\
$^{3}$Department of Astronomy, University of Florida, 211 Bryant Space Science Center, Gainesville, FL 32604, USA 
}

\begin{document}

%\date{Accepted 1988 December 15. Received 1988 December 14; in original form 1988 October 11}

\pagerange{\pageref{firstpage}--\pageref{lastpage}} \pubyear{2006}

\maketitle

\label{firstpage}

\begin{abstract}

We have used the Very Large Telescope (VLT) to obtain spectropolarimetry of 
the radio-loud, double-lobed broad absorption line (BAL) quasar PKS 0040$-$005.
We find that the optical continuum of PKS 0040$-$005 is intrinsically 
polarized at 0.7\%  with an electric vector position angle nearly parallel
to that of the large-scale radio axis.  This result is naturally explained 
in terms of an equatorial scattering region seen at a small inclination, 
building a strong case that the BAL outflow is {\em not} equatorial.   
In conjunction with other recent results concerning radio-loud BAL quasars, 
the era of simply characterizing these sources as ``edge-on'' is over.

\end{abstract}


\section{Introduction}

Depending on the selection techniques used, some 10-30\% of luminous
quasars show blueshifted broad absorption lines (BALs) indicative of 
substantial outflows.  By carrying away angular momentum, these outflows 
may be a fundamental part of the accretion process, and they may also
be important for chemically enriching the interstellar and even intergalactic
mediums.  There is still no consensus about the nature of BAL quasars,
however, although many possibilities have been discussed 
(e.g. Weymann et al. 1991).  

One possibility is that quasars only possess such high-velocity outflows 
during a relatively short-lived evolutionary phase during which they blow 
material out of the nuclear region (e.g., Voit et al. 1993; Becker et al. 
2000; Gregg et al. 2002; Gregg et al. 2006).   Another is that BAL quasars 
and normal quasars may be unified through orientation as for some other 
AGN classes.  In this popular equatorial paradigm,
BALs are seen in quasars viewed at high inclination angles such that the
line of sight passes through an equatorial wind (e.g., Hines \& Wills 1995; 
Cohen et al. 1995; Goodrich \& Miller 1995; Murray \& Chiang 1995).

Polarimetry can be a useful tool for testing the latter idea, ``unification
by orientation,'' because scattering processes leading to polarization when 
geometric asymmetries are present.  Antonucci (1983) first pointed out that
Seyfert 1 and Seyfert 2 galaxies both display optical polarization,
but in the type 1 galaxies the polarization electric vector is parallel
to the radio axis, while in type 2 galaxies the two are perpendicular,
suggesting different geometries for the two types and helping to lead
to the unification of Seyfert galaxies through orientation.
Spectropolarimetry of NGC 1068 (Antonucci \& Miller 1983) showing a 
type 1 broad-lined spectrum in polarized light made the case compelling.
Broad-lines and continuum may be scattered by a polar scattering region
along the jet axis, leading to perpendicular angles, while equatorial
scattering in a Seyfert 1 leads to the parallel angles.  This work has
been greatly expanded in recent years (e.g., Smith et al. 2004) and
also applied to radio-loud AGNs at both low and high redshift
(e.g., Cimatti et al. 1993; Cohen et al. 1999; Vernet et al. 2001).

\begin{figure*}
\begin{center}
\includegraphics[width=11.5 truecm]{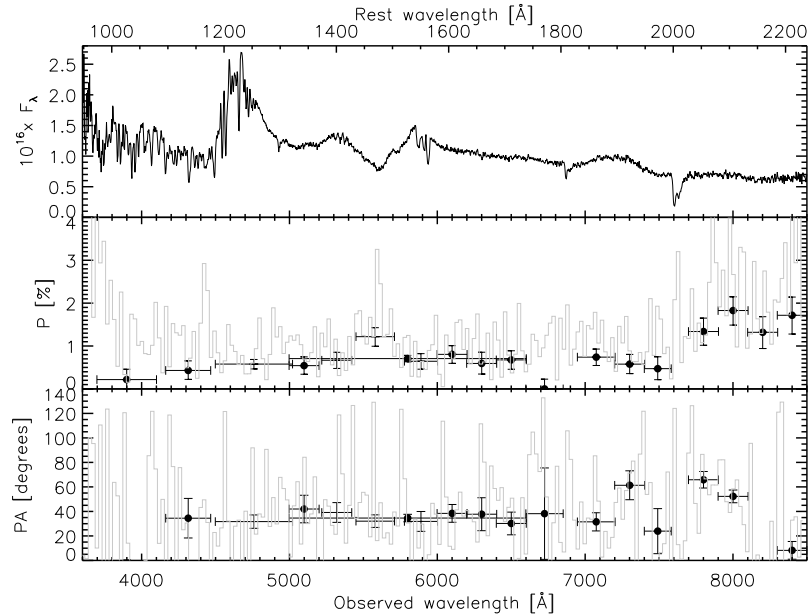}
\end{center}
\caption{Spectropolarimetric results for PKS$-$005.  
The top panel shows the total flux spectrum in units of ergs s$^{-1}$ cm$^{-2}$ \AA$^{-1} \times 10^{16}$. 
The middle panel shows the debiased percentage linear polarization,
while the bottom panel shows the position angle.  Error bars on the big
bins represent 1 $\sigma$ uncertainties.
The light grey bins in the bottom panels are 9 \AA.}
\end{figure*}

When radio-loud BAL quasars began to be found 
(Becker et al. 1997; Brotherton et al. 1998; Becker et al. 2000), 
Brotherton et al. (1997) proposed that the alignments of the jet axes
and polarization angles should be measured to test BAL quasar geometry in the
same way as in other AGNs.  Despite the discovery of radio-loud BAL quasars,
very few have shown extended jets permitting this test.

In order to pursue questions concerning the nature of BAL quasars,
we searched for BAL quasars with extended radio structures by examining
Sloan Digital Sky Survey quasars (Schneider et al. 2002) and their radio maps 
from FIRST Survey (Becker, White, \& Helfand 1995).  
We classified SDSS~004323.43$-$001552.43 as an extended, radio-loud
FR II BAL quasar.  This quasar was previously
detected by many radio surveys and is better known as PKS 0040$-$005.
PKS 0040$-$005 ($z = 2.806$) has also been recently classified as a FR II 
BAL quasar by Gregg et al. (2006) and Zhou et al. (2006).  We targeted 
PKS 0040$-$005 for spectropolarimetry to test the geometry of BAL quasars.  

\section[]{Observations and Data}

\subsection{VLT Spectropolarimetry}

We obtained spectropolarimetric observations of
PKS 0040$-$005 on UT 2003 September 20 and 21 using the PMOS
mode of the FORS1 spectrograph \citep{app98} on the Antu unit of the
ESO Very Large Telescope (VLT). On September 20, we observed the source at
airmass between 1.83 and 1.62 with 1.0\arcsec\ seeing. To obtain more
sensitive data, we re-observed the source during the next night with
airmass 1.15 and 0.7\arcsec\ seeing. Conditions were photometric on
both nights. The observations were split into four 300s exposures,
each at a different orientation of the half-wave plate (0$^{\circ}$
22.5$^{\circ}$, 45$^{\circ}$, 67.5$^{\circ}$). We used the 300V grism 
with a 1.0\arcsec\ wide slit oriented North-South, 
resulting in a spectral resolution of $\sim$10~\AA (FWHM).

We reduced the data using standard IRAF procedures. We extracted
the spectra with identical 4\arcsec\ wide apertures for the o and
e-rays, and re-sampled all 8 individual spectra (o and e-rays for the
4 half-wave plate positions) to the same linear dispersion
(2.6\AA/pix) in order to calculate the polarization in identical
spectral bins. We used the procedures of Vernet (2001; see also Vernet
et al. 2001)\nocite{ver01a,ver01c}, which are based on the method
described by Cohen et al. (1997). We checked the polarization angle offset
between the half-wave plate coordinate and the sky coordinates against
values obtained for the polarized standard stars BD\_12\_5133 and
NGC~2024~NIR1 and the unpolarized standard star GD~50; our values are
within $<1^{\circ}$ from the published values, and the polarization
percentage within 0.1\%. Figure~1 shows our results.

\subsection{Supplemental Data and Information}

The 1.5GHz map of PKS $0040-005$ from FIRST (Becker et al. 1995) shows
a double-lobed source that has a position angle of 51$^{\circ}$.
Gregg et al. (2006) show the map, and provide additional measurements based 
on the radio image and the optical SDSS spectrum.  They assume a cosmology
of $H_0= 71$\ km s$^{-1}$\ Mpc$^{-1}$\ and $q_0 = 0.5$, and report a
1.4 GHz radio luminosity (ergs s$^{-1}$) of log $L_{1.4 GHz}$ = 34.1,
an absolute $g$ magnitude of $M_g = - 26.44$ (and $-$26.78 after 
correcting for an estimate of small intrinsic reddening), and a 
radio loudness parameter log R* (Stocke et al. 1992) of 2.9 (2.7 dereddened).
The radio source is 117 kpc across for this cosmology,
the smallest of their eight sources.  The radio core dominance,
or core-to-lobe flux at 1.4 GHz, they report to be 0.09.  The C IV
BAL has a balnicity index (Weymann et al. 1991) of 1073 km s$^{-1}$.

The homogenized radio data available through the NASA Extragalactic 
Database (NED) 
shows that the radio spectrum of PKS $0040-005$ is steep, with a
radio spectral index $\alpha$ = $-1.2$ ($S_{\nu} \propto \nu^{\alpha}$).
 
\section{Results}

Averaging over the central spectral region where the signal-to-noise
ratio is high and the data are well behaved, PKS 0040-005 shows 
intrinsic continuum polarization ($0.72\pm0.08\%$) at a well-defined position 
angle($36\pm3^{\circ}$).   The polarization is intrinsic.  The interstellar
polarization (ISP) from aligned dust grains is expected to be very low 
toward this sight line, since the Schlegel et al. (1998) extinction E(B-V) 
is only 0.017 mag, and the maximum ISP polarization should be less than
$9 \times E(B-V) \%$ (Serkowski et al. 1975), or 0.15\%.  
Furthermore, the polarization rises in the C IV BAL trough bottom 
(and perhaps Lyman $\alpha$) showing that the polarization 
must be intrinsic.  Rising polarization in trough bottoms is seen 
in many BAL quasars and has been interpreted as the result of the 
scattered line of sight being less absorbed than the direct line of 
sight (e.g., Ogle et al. 1999).

While we prefer an explanation of scattering as the origin of the polarization,
the data quality is insufficient to completely rule out synchrotron emission.  
The polarized flux spectrum is noisy and not inconsistent 
with a power-law.  PKS 0040$-$005 is a steep spectrum FR II radio source with
strong lobes and a weak core, suggestive that it is not being viewed directly
jet on which is usually required for synchrotron to be important, but 
the polarization level here is relatively weak and it is possible and 
has been seen in other sources (Schmidt \& Smith 2000).  We argue below, under
the assumption that the polarization arises from scattering, that 
PKS 0040$-$005 and its BAL outflow is not seen at a particularly high
inclination angle.  The case for this is only stronger if the polarization
is from beamed synchrotron.

\section{Discussion}

The position angle of the continuum polarization ($36\pm3^{\circ}$) is 
close to parallel to the large-scale radio axis ($\sim51^{\circ}$), a difference
of 15$^{\circ}$, and is the first BAL quasar found with this property.  
In analogy to the Seyfert galaxies and radio galaxies/quasars,
this suggests a geometry in which PKS 0040$-$005 is not seen edge-on, but
at some modest inclination angle (consistent with the smallish but double-lobed
radio structure), and the scattering takes place in an
equatorial scattering region.  Good illustrations of this geometry can
be seen in Figures 6 and 10 of Smith et al. (2004).  

Lamy \& Hutsem{\'e}kers (2004) explored the concept of a ``two component wind''
for BAL quasar outflows, that included both equatorial and polar
outflows, and had support from theoretical models (e.g., Proga 2000, 2003;
Pereyra et al. 2004), which are variations of the disk wind of 
Murray et al. (1995).  Lamy \& Hutsem{\'e}kers (2004) were trying to 
explain not only some correlations they found in the polarimetric properties
and their relationship to BAL properties, but also the source of the 
polar scattering region.  The fact that sometimes polarization angle rotations 
are seen across emission lines in BAL quasars, or with changes in wavelength,
is suggestive that more than one scattering path is likely present.
Our observation of PKS $0040-005$ fits into this picture, and in fact its
other properties are also consistent with the relationships 
Lamy \& Hutsem{\'e}kers (2004) report: the continuum polarization is low
which corresponds to a large BAL detachment and weak absorption in the
polarized flux spectrum.

We can compare the alignment in PKS $0040-005$ to other BAL quasars.
Table 1 summarizes the optical polarization properties and radio position
angles of BAL quasars for which both quantities have been reported.
References provided for each are relevant to the discovery or classification,
or provide details for the information given in the table.  Some entries
are averages over a wide range of wavelengths (e.g., broad band or white
light observations), but when explicit ranges
are given, it indicates that the values change as a function of wavelength.

\begin{table*}
 \centering
 \begin{minipage}{140mm}
  \caption{Polarization-Radio Alignments of Broad Absorption Line Quasars}
  \begin{tabular}{@{}lllccccl@{}}
  \hline
   Object & $z$ &Type   & \multicolumn{2}{c}{Optical Polarization} & Radio P.A. & Angular & References \\ 
   &  & & Pol. (\%)& $\theta_P$ ($^{\circ}$) & $\theta_R$ ($^{\circ}$) & $\Delta \theta$ ($^{\circ}$)& \\
%   & &  & &  &  &   \\
\hline
PKS 0040$-$005 & 2.81 & HiBAL & $0.72\pm0.08\%$ & $36\pm3$ & 51 & 15 & This Paper \\
 & &  & & &  &  &   \\
FIRST J1016+5209 & 2.46 & HiBAL & 2-3 & 75-85 & 146 & 71-81 & Gregg et al. 2000 \\
UN J1053$-$0058 & 1.55 & LoBAL & 1.9 & 90 & 27 & 63 & Brotherton et al. 1998\\
 & & & & & & & Lamy \& Hutsem{\'e}kers 2000 (pol) \\
FBQS 1312+2319 & 1.52 &HiBAL & 1.1 & 166 & 59 & 83 & Becker et al. 2000 \\
& & & & & & & Sluse et al. 2004 (pol)\\
& & & & & & & Jiang \& Wang 2003 (rad)\\
LBQS 1138$-$0126 & 1.27 & LoBAL & 3-4 & 150-180 & 52 & 52-82 & Brotherton et al.
 2002 \\
PG 1700+5153 & 0.29 & LoBAL & 0.6 & 55 & 145 & 88 & Kellerman et al. 1994 (rad)  \\
 &  & & &  &  & & Schmidt \& Hines 1999 (pol) \\
PKS 1004+130 & 0.24 & HiBAL & 0.7-1.6 & 24-63 & 117 & 54-87 & Wills et al. 1999 \\
& & & & & & & Webb et al. 1993 (pol) \\
%Mrk 231 & & LoBAL & 3-15 & 93 & 65 & 27 & Smith et al. 1995 \\
%    &  & & & &  & &  Ulvestad et al. 1999 (rad) \\
\hline
\end{tabular}
\end{minipage}
\end{table*}

\begin{figure}
\begin{center}
\includegraphics[width=8.5 truecm]{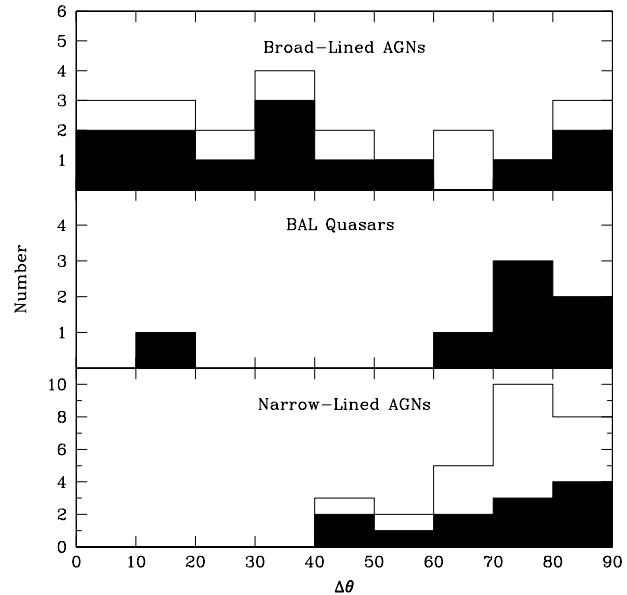}
\end{center}
\caption{Histograms of $\Delta \theta$, the difference between the 
radio axis position angle and the electric vector of the optical/ultraviolet
continuum linear polarization. The top panel shows broad-lined type 1 AGNs,
including Seyfert 1 galaxies (shaded), broad line radio galaxies, and some 
scattered-light quasars (see text).  
The bottom panel shows narrow-lined type 2 AGNs, including hidden-broad-line 
Seyfert 2 (shaded) and radio galaxies (see text).  
The middle panel shows BAL quasars from Table 1.}
\end{figure}

Figure 2 shows histograms of the difference in position angle between
the electric vector of the continuum polarization and the radio axis
for BAL quasars from Table 1, and for comparison also broad-lined type 1 
and narrow-lined type 2 AGNs.  When the position angle has a range,
we take the middle value to place the object in a bin.  
The broad-line AGNs include 
Seyfert 1 galaxies from Brindle et al. (1990), broad line radio galaxies
from Cohen et al. (1999), and several higher redshift radio-loud quasars with
high scattering polarization (Goodrich \& Miller 1988;
Brotherton et al. 1998; De Breuck et al. 1998). 
Type 1 AGNs with scattering polarization, both radio loud and radio quiet, 
show both parallel and perpendicular behavior.  The type 2 AGNs
include Seyfert 2 galaxies from Brindle et al. (1990), using only those
with hidden broad-line regions (see Tran 2003) and well-defined angles, 
and high-redshift radio galaxies from Vernet et al. (2001).  
When type 2 objects are restricted to those with scattering polarization, 
the radio and polarization axes are perpendicular.

Before our observation of PKS 0040$-$005, it would have been easy to
conclude based only on the available angle differences that BAL quasars
were, like type 2 AGNs, seen edge-on.  There are some selection effects
at work, however, with so few BAL quasars having this measurement 
available.  A ``jet on'' quasar will be a smaller radio source than
it would with the jet in the plane of the sky (as would young sources), 
and will be less
likely to be resolved in FIRST maps.  Furthermore, sources dominated by 
equatorial rather than polar scattering often have weak polarization 
due to larger dilutions from the face-on quasar (although dust present
in some BAL outflows, especially low-ionization BAL quasars, may also
reduce dilution), and for AGNs seen directly face-on the scattering 
polarization is expected to be zero.  For instance, the highly polarized
low-ionization BAL quasar FIRST J1556+3517 (Brotherton et al. 1997)
has a flat and variable radio spectrum, signs of a jet-on source,
but has not been resolved even with VLBI (Jiang \& Wang 2003).

While our spectropolarimetry of PKS 0040$-$005 represents only one data point,
it joins a growing list of observations that pose problems for the 
equatorial paradigm.
 
\subsection{More Evidence Against Edge-On Only Orientation in BAL Quasars}

The present observation is not the only evidence against simple orientation
schemes.  Over the last decade, evidence against edge-on geometries has
emerged from several avenues, the strongest involving radio emission.  
The properties of radio sources have long been understood to depend greatly 
upon how they are seen.

Barvainis \& Lonsdale (1997) and Becker et al. (2000) pointed out that
BAL quasars have both flat and steep radio spectra.  Typically, jet-on
sources with optically thick beamed synchrotron radiation have flat 
radio spectra, while edge-on sources are dominated by optically thin
radio lobe emission which has a steep spectrum.  The first study was on
formally radio-quiet objects, but the Becker et al. (2000) sample has
radio-loud objects of both types.  Furthermore Becker et al. (2000) noted
that very few of the radio-selected BAL quasars were extended at FIRST Survey
resolutions compared to similarly selected, unabsorbed quasars.  Some
of these BAL quasars were compact steep spectrum sources, which have been
hypothesized to be young sources (O'Dea 1998).  Gregg et al. (2002)
and Gregg et al. (2006), studying examples of FR II BAL quasars developed
this idea into an evolutionary hypothesis, that a BAL phase evolves into
a radio-loud phase, with only a short temporal overlap.  Zhou et al. (2006)
find a number of BAL quasars associated with radio sources that vary
on relatively short timescales indicating extreme brightness temperatures
that can only be explained by beaming within 20 degrees of the jet direction.
Another radio oddity for BAL quasars is noted by White et al. (2006),
reporting that BAL quasars on average seem to have higher average radio 
fluxes than unabsorbed quasars (it should be noted that the majority
are not radio loud).  These radio properties seem quite difficult
to explain in terms of a single preferred orientation. 

Additionally, at optical/ultraviolet wavelengths, BAL quasars have been noted 
to lie at an extreme of the ``eigenvector 1'' of Boroson \& Green (1992), 
showing an excess of broad optical Fe II emission and a deficit of [O III] 
emission (Boroson 2002; Yuan \& Wills 2002).
Turnshek et al. (1997) followed up on this idea, using the Hubble Space 
Telescope (HST) to obtain ultraviolet spectra of quasars with weak [O III]
emission, finding approximately 1/3 of these to show C IV BALs. 
Telfer et al. (2000) conducted an HST imaging survey 
for extended narrow-line regions they expected to see if BAL quasars
were edge-on quasars, but no such extensions were found.
The weak NLR emission in particular is difficult for orientation models to 
explain, since the spectra of edge-on type 2 objects are remarkable for their 
prominent narrow emission lines. 

X-ray observations of BAL quasars may also be problematic for the 
equatorial paradigm.  Punsly (2006) points out that the absorbing columns
measured with X-ray telescopes toward BAL quasars are not smaller than those
toward type 2 quasars.  Disk wind models predict they
should be, and Punsly argues that the observational selection effects 
seem insufficient to account for this inconsistency.

Finally, we have omitted one object from Table 1 that may also have aligned axes
and a polar BAL outflow, but the situation is not clear cut.  Markarian 231 has 
appeared in the literature with both small (Antonucci 1983) and large 
$\Delta \theta$ (Smith et al. 2004).  The VLBA observations of Ulvestad et al. 
(1999) may indicate a jet direction ($\sim$65$^{\circ}$) quite different from 
the large scale radio structure ($\sim$5$^{\circ}$), which may be an issue 
of general concern, although Mrk 231 is not radio-loud like the high-redshift
entries in the table.  Another difficulty involves the fact that the 
polarization angle rotates significantly in the far ultraviolet 
(Smith et al. 1995), and that in Mrk 231 and other objects it is not always 
clear at what wavelength the polarization position angle should be compared
to the radio position angle.  Angle rotations indicate multiple scattering
axes in a number of BAL quasars, and type 1 Seyferts can show both polar
and equatorial scattering (Smith et al. 2004).  Supporting a polar outflow
in Mrk 231, Punsly \& Lipari (2005) claim there is an 
outflow aligned with the small-scale jet angle 
based on two-dimensional spectroscopy of the blue wing of H$\alpha$.

\section{Conclusions}

PKS 0040$-$005 is a radio-loud BAL quasar with low but intrinsic continuum
polarization.  The electric vector of the continuum polarization 
is close to parallel to the position angle of the quasar's extended radio
structure.  This is naturally explained as the result of scattering in an
equatorial region with a face-on geometry, just as in Seyfert 1 and
broad-line radio galaxies.  Our result, taken in conjunction with other
results in the literature, seems to be the end of the popular but 
apparently too simple idea that BAL outflows are always equatorial. 

\section*{Acknowledgments}

We thank Joel Vernet for his reduction
software package and discussions concerning the data.  MSB 
thanks the ESO Scientific Visitor Programme for support and hospitality.
This research has made use of the NASA/IPAC Extragalactic Database (NED) 
which is operated by the Jet Propulsion Laboratory, California Institute of 
Technology, under contract with the National Aeronautics and Space 
Administration.  
%NSF acknowledgment?  
%We also thank radioactive Cornholio for not messing up the observations.

\label{lastpage}

\end{document}